\documentstyle[11pt,newpasp,twoside,epsf]{article}
\def\ts{\thinspace}
\def\edcomment#1{\iffalse\marginpar{\raggedright\sl#1\/}\else\relax\fi}
\reversemarginpar

\begin{document}
\title{The Nature of the faint far-infrared extragalactic source
       population: Optical/NIR and radio follow-up observations of
       ISOPHOT deep-field sources using Keck, Subaru, and VLA telescopes}

\author{Yuko Kakazu$^{1}$, 
D. B. Sanders$^{1}$, 
R. D. Joseph$^{1}$, 
L. L. Cowie$^{1}$,
T. Murayama$^{2}$,
Y. Taniguchi$^{2}$,
S. Veilleux$^{3}$,
M. S. Yun$^{4}$,
K. Kawara$^{5}$,
Y. Sofue$^{5}$,
Y. Sato$^{5}$,
H. Okuda$^{6}$,
K. Wakamatsu$^{7}$,
T. Matsumoto$^{8}$,
and H. Matsuhara$^{8}$}

\affil{\footnotesize
$^{1}$Institute for Astronomy, University of Hawaii.
$^{2}$Astronomical Institute, Tohoku Univesity.
$^{3}$Deptartment of Astronomy, University of Maryland.
$^{4}$FCRAO, University of Massachusetts.
$^{5}$Institute of Astronomy, University of Tokyo.
$^{6}$Gumma Observatory.
$^{7}$Gifu University.
$^{8}$ISAS.}

\begin{abstract}
We report on optical and near-infrared (NIR) follow-up spectroscopy
of faint far-infrared (FIR) sources found in our deep FIR survey
by Kawara et al. 
\end{abstract}

\section{Introduction}
Deep surveys at FIR and submilimeter wavelengths have been carried out in
order to investigate the nature of dust-enshrouded galaxies at high
redshift. As a contribution to this field, our group made a deep FIR
survey using the ISOPHOT camera on board the {\it Infrared Space
Observatory} (ISO) satellite (Kawara et al. 1998; Matsuhara et al. 2000).
Mapping at 90$\mu$m and 170$\mu$m of two $44^\prime \times 44^\prime$
fields in the Lockman Hole (LH\_EX and LH\_NW), a region exhibiting the
lowest H{\ts}{\sc i} column density in the sky (Lockman et al. 1986),
resulted in the detection of 36 sources with $f_{90} >${\ts}150{\ts}mJy
and 45 sources with $f_{170} >${\ts}150{\ts}mJy.  Given the relatively
large size of the ISOPHOT beam at 170$\mu$m ($\sim$90$^{\prime\prime}$), 
we have obtained opt/NIR images and spectra using telescopes on Mauna Kea
and 6cm radio continuum maps using the VLA (Yun et al. 2002) to identify
the most likely source of the 170$\mu$m emission.  Here we report our
initial identifications of the brightest of the ISOPHOT 170$\mu$m
sources. 

\section{Results and Discussion}
Redshifts of 35 FIR source candidates were determined using optical
spectra obtained with ESI on Keck{\ts}II during three observing runs
in 2000 March and 2001 January.  Infrared luminosities, $L_{\rm
ir}(8-1000\mu m)$, were then estimated by using the ISOPHOT fluxes and
assuming an SED similar to that of Arp{\ts}220.  We found one
hyperluminous infrared galaxy (HyLIG: $L_{\rm ir} > 10^{13} L_\odot$) at
$z=1.6$, 11 ultraluminous infrared galaxies (ULIGs: $L_{\rm ir} > 10^{12}
L_\odot$) at $0.3 < z < 1$, 12 luminous infrared galaxies (LIGs:
$L_{\rm ir} > 10^{11} L_\odot$), and 11 galaxies with $L_{\rm ir} <
10^{11} L_\odot$. Except for one LIG at $z=0.365$, all of the galaxies
with $L_{\rm ir} < 10^{12} L_\odot$) are at $z<0.3$.  The mean redshift
for all sources is $0.31\pm0.31$.

The low-resolution ESI spectra were used to determine the optical
spectral-type of the candidate ISOPHOT sources. Following procedures
used by Murayama \& Taniguchi (1998), the spectra were classified into
four types -- AGNs, LINERs, HII-type, and early-type (without
emission lines).  The HyLIG at $z=1.6$ was found to be a
quasar.  One ULIG had an early-type spectrum and 10 ULIGs are HII
galaxies.  Among the remaining 23 lower-luminosity sources, there was one
early-type galaxy, one Seyfert 2, 10 LINERs and 11 HII galaxies.
Thus, based on our low-resolution ESI optical spectra most of the ISOPHOT
175$\mu$m sources appear to be powered primarily by star formation,
consistent with the conclusion reached from an analysis of ISOPHOT
number counts by Matsuhara et al. (2000) that most of the ISOPHOT sources
are star-forming galaxies at $z<1$.

\begin{figure}
\plotfiddle{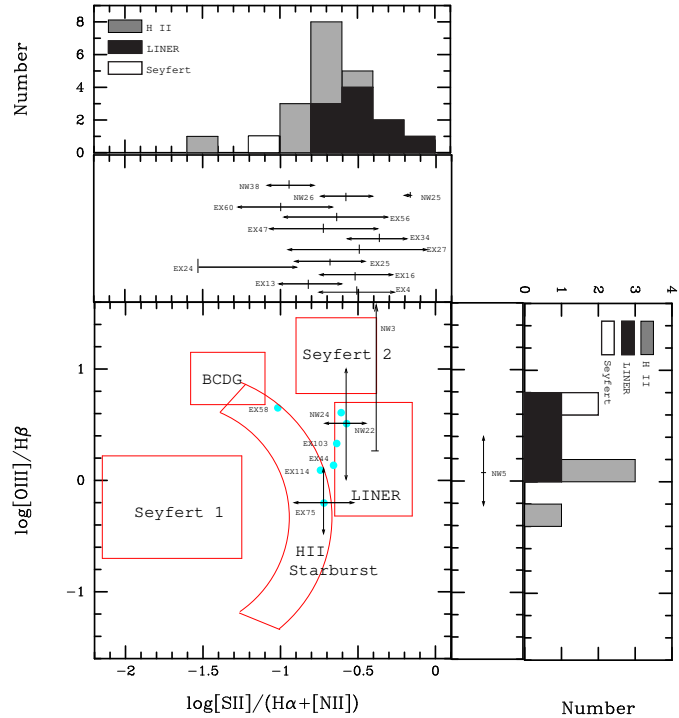}{250pt}{0}{50}{50}{-150}{-68}
\caption{Low-resolution emission line diagostics of ISOPHOT source
candidates.}
\end{figure}


\vskip 0.1cm
\begin{references}
\vskip -0.1cm
\reference{Kawara, K., et al. 1998, A\&A, 336, L9}
\reference{Lockman, F. J., Jahoda, K. \& McCammon, D. 1986, ApJ, 302, 432}
\reference{Matsuhara, H., et al. 2000, A\&A, 361, 407}
\reference{Murayama, T. \& Taniguchi, Y. 1998, PASJ, 50,241} 
\reference{Yun, M. S., et al. 2002, in preparation}
\end{references}
\end{document}